\documentclass[
]{ceurart}

\usepackage{listings}
\usepackage{float}
\usepackage{wrapfig}
\usepackage{mdframed} 
\usepackage{enumitem} 
\begin{document}

\newcommand{\TODO}[1]{\textcolor{red}{\begingroup\raggedright TO DO: #1\\\endgroup}}
\newcommand{\NOTE}[2][gray]{\smallskip\noindent
  \colorbox{#1!30}{\parbox{.98\linewidth}{{\small\textbf{#2}}}}
}

\copyrightyear{2025}
\copyrightclause{Copyright for this paper by its authors.
  Use permitted under Creative Commons License Attribution 4.0
  International (CC BY 4.0).}

\conference{AI Literacy for All}

\title{AI Literacy Assessment Revisited: A Task-Oriented Approach Aligned with Real-world Occupations}


\author{Christopher Bogart}[%
orcid=0000-0001-8581-115X,
email=cbogart@andrew.cmu.edu
]

\address{Carnegie Mellon University,
  Pittsburgh, PA, USA}

\author{Aparna Warrier}[%
orcid=0009-0004-2502-847X ,
email=aparnamw@andrew.cmu.edu,
]

\author{Arav Agarwal}[%
orcid=0000-0001-9848-1663,
email=arava@andrew.cmu.edu,
]

\author{Ross Higashi}[%
email=rhigashi@andrew.cmu.edu,
]

\author{Yufan Zhang}[%
orcid=0009-0004-3791-7311,
email=yufanz@andrew.cmu.edu,
]

\author{Jesse Flot}[%
email=jbflot@nrec.ri.cmu.edu,
]

\author{Jaromir Savelka}[%
orcid=0000-0002-3674-5456,
email=jsavelka@andrew.cmu.edu,
]

\author{Heather Burte}[%
orcid=0000-0000-0000-0000,
email=hburte@andrew.cmu.edu,
]

\author{Majd Sakr}[%
orcid=0000-0001-5150-8259,
email=msakr@andrew.cmu.edu,
]


\begin{abstract}
As artificial intelligence (AI) systems become ubiquitous in professional contexts, there is an urgent need to equip workers, often with backgrounds outside of STEM, with the skills to use these tools effectively as well as responsibly, that is, to be AI literate. However, prevailing definitions and therefore assessments of AI literacy often emphasize foundational technical knowledge, such as programming, mathematics, and statistics, over practical knowledge such as interpreting model outputs, selecting tools, or identifying ethical concerns. This leaves a noticeable gap in assessing someone's AI literacy for real-world job use. We propose a work-task-oriented assessment model for AI literacy which is grounded in the competencies required for effective use of AI tools in professional settings. We describe the development of a novel AI literacy assessment instrument, and accompanying formative assessments, in the context of a  US Navy robotics training program. The program included training in robotics and AI literacy, as well as a competition with practical tasks and a multiple choice scenario task meant to simulate  use of AI in a job setting. We found that, as a measure of applied AI literacy, the competition's scenario task outperformed the tests we adopted from past research or developed ourselves. We argue that when training people for AI-related work, educators should consider evaluating them with instruments that emphasize highly contextualized practical skills rather than abstract technical knowledge, especially when preparing workers without technical backgrounds for AI-integrated roles.
\end{abstract}


\maketitle

\section{Introduction}

As artificial intelligence (AI) systems become increasingly integrated into the workplace, there is a growing need to equip professionals, often with backgrounds outside of STEM, with the skills to use these tools effectively and responsibly. Many roles now entail interaction with AI, not in the sense of developing or programming systems, but in choosing when and how to use them, and knowing how to appropriately use or question their outputs. However, some definitions and assessments of AI literacy overemphasize technical implementation details, specialized knowledge, and concepts from programming, mathematics, and statistics, rather than emphasizing applied, domain-specific interaction. 

In this paper, we adapt AI literacy definitions to emphasize workplace-relevant competencies over abstract technical knowledge. We explore this model in the context of a U.S. Navy training program for robotics warfare specialists---a role that requires interaction with AI systems but not their development. This setting exemplifies the need for tailored ``AI user'' training and assessments.

We describe the development of a novel AI literacy assessment instrument, beginning with an adaptation of a prior measure by \citet{hornberger_what_2023}. After piloting this version with the program’s first cohort, we conducted an item-level analysis and substantially revised the instrument for the second cohort. Alongside these assessments, both cohorts participated in a series of applied training modules focused on practical uses of AI. The second cohort concluded with a competition modeled on realistic job tasks, including scenario-based multiple choice questions involving AI-relevant decision-making, as well as tasks such as selecting and labeling training data for an AI model used in a robotics context.

While performance on these applied tasks improved following training, gains were not reflected in the overall AI literacy scores on either version of the generic AI literacy instrument, nor did the instrument broadly predict performance on competition tasks. However, one of the competition tasks, a multiple-choice scenario-based questionnaire turned out to be highly correlated with some of the remaining tasks, suggesting a new way of designing AI literacy assessments.

This paper investigates that discrepancy by analyzing item-level patterns and exploring possible mismatches between existing AI literacy measurement methods and practical competence. Though limited in scope and sample size, our findings underscore the need to ground both AI training and assessment in authentic work tasks, and to reconsider how AI literacy is defined and measured in professional contexts.

\subsection{Research Questions}
\begin{enumerate}

    \item \textbf{To what extent do AI literacy scores reflect the effects of AI literacy training?} \\
    We examine both overall scores and item-level patterns across two versions of our AI literacy instrument, plus a competition task reframed as an AI literacy assessment, to assess alignment with observed task competence.

    \item \textbf{To what extent do AI literacy scores predict performance on AI-relevant tasks?}

    \item \textbf{What features of the AI literacy instruments make them effective or ineffective as measurements of training or predictors of task performance?}
    We consider components of the AI literacy as well as the questions' Bloom taxonomy level and domain-specificity.
\end{enumerate}

\section{Related work}

AI literacy has been defined in various disciplines \cite{carolus2023,Long2020WhatIA,hermann2021}, spanning human-computer interaction, computer education, and information systems \cite{pinski2024}. AI education itself ranges from university courses to youth programs \cite{williams_ai_2023,touretzky_envisioning_2019} and military training \cite{salazar-gomez_designing_2022}. Although these educational efforts differ in their perspectives---targeted at different audiences as well as learning outcomes \cite{chiang2022,leichtmann2023}---they revolve around a set of proficiency dimensions in subject areas \cite{pinski2024} that are stabilizing in the literature. Distinctions between AI literacy and other technology-related literacies such as scientific literacy \cite{laugksch2000}, digital literacy \cite{eshet2024}, and media literacy \cite{potter2004} are also being drawn, highlighting the unique characteristics of learnability, autonomy, and inscrutability of AI as a technology \cite{berente2021}.

Instruments for measuring AI literacy have adapted accordingly, some for specific user groups \cite{berente2021} and courses \cite{perchik_artificial_2023}, others more generalizable. While most instruments are based on self-assessments \cite{wang2022,laupichler2023,karaca2021,carolus2023} which are subject to individual biases \cite{chiu2024}, some measure changes in performance \cite{lintner2024}. Kong et al. devised an AI concepts test consisting of 7 multiple choice questions (MCQs), combined with 4 additional surveys to measure changes in self-assessments \cite{kong2024}. Zhang et al. devised an  AI literacy concept inventory (AI-CI) assessment consisting of 20 MCQs in the topic areas of ``AI general concepts,'' ``logic systems,'' ``machine learning general concepts,'' and ``supervised learning'' \cite{zhang2024}. Sanfiel et al. devised a scale of AI literacy for all (SAIL4ALL), adapting Long \& Magerko's framework, consisting of 52 confidence scale questions based on ``What is AI?,'' ``What can AI do?,'' ``How does AI work?,'' ``How Should AI Be Used?'' \cite{sanfiel2024}. Markus et al. combined validated instruments from a dozen sources while supplementing with their own \cite{markus_objective_2025}.

\section{AI Literacy instrument design approach}

Our approach is grounded in the belief that for AI Literacy to be a useful measure of one's ability to navigate an AI-enabled world effectively, it need not measure understanding of abstract AI concepts; rather it should measure the ability to engage practically with AI in civic, consumer, and workplace contexts. Since we are focused on workplace preparation, we are designing the instrument to focus on workplace skills---knowing when and how to use AI tools to enhance job performance and when to critically assess AI outputs. As such, the instrument should focus on real-world interactions with AI systems, reflecting the challenges that professionals face when using AI in their everyday work.

This is in contrast to more traditional, theory-heavy approaches to AI literacy, which often emphasize technical knowledge in isolation. We aim to design an instrument that measures the skills needed for AI-human collaboration, trust calibration, and troubleshooting. Our AI User training modules similarly emphasize contextualized, hands-on learning. Together the training and instrument support the development of practical AI literacy that is both usable and useful in a professional setting.

We discuss three proposed AI literacy instruments, developed as follows:
\begin{itemize}
    \item \textbf{Hornberger (AI-LIT-H)} In our first iteration, we used a mostly unmodified version of Hornberger's~\cite{hornberger_what_2023} AI literacy test. This instrument was chosen for its comprehensive approach to assessing
    non-expert understanding of AI; Hornberger designed it by winnowing down the larger set of categories proposed by Long and Magerko~\cite{long_what_2020}. We omitted only one of Hornberger's 31 questions, Q18, for ease of implementation because it was the only non-multiple choice question.

    \begin{mdframed}[backgroundcolor=gray!10, linewidth=0.5pt]
    \textbf{Example question:}\\
    Which of the following interdisciplinary research fields is also a subfield of AI?\\
    A. Natural Language Processing \textbf{(correct answer)}\\
    B. Blockchain\\
    C. Psychology of Learning\\
    D. Bioinformatics
    \end{mdframed}

    \item \textbf{Modified Hornberger (AI-LIT-MH):}
    After an analysis of the first cohort's results (see Section~\ref{sec:ailit_item_analysis}), we dropped 17 questions that we deemed esoteric, outdated, or off topic. For example, one of Hornberger et al's original questions suggests a human being could tell if they are interacting with another human being or an AI agent by making ironic remarks, which were better understood by humans than by AI, but since 2022, AI has evolved to understand ironic remarks quite well. We then generated 13 additional questions to replace missing competencies or add new ones.

Our AI Literacy test includes nine categories. We omitted several of Long and Magerko’s original categories that we judged to have limited relevance to practical workplace use. These included philosophical or academic concepts such as Understanding Intelligence and General vs. Narrow AI, broad interdisciplinary knowledge, robotics-specific topics like Sensors and Action \& Reaction, and overlapping categories such as Decision-Making, which we subsumed under more practical competencies like data and performance evaluation.

We also added new categories aligned with the hands-on AI User training, which were missing from previous frameworks: Evaluating AI Outputs and Evaluating Model Performance (splitting Long and Magerko’s “Critically Interpreting Data”), Matching AI Tools to Tasks, and Collaboration with AI Systems, reflecting workplace decisions about when and how to use AI in real-world settings.

    The second cohort used this revised test, and we omitted three of its questions from the analysis for which we later decided correct answer choice was not unambiguously correct: two, \texttt{Limit2} and \texttt{Limit3}, from the original Hornberger test, and one, \texttt{Output1}, which was our own. Although we did not have enough participants to perform a rigorous item analysis, we used the analysis to reconsider face validity and removed these three.

    It is important to note that the \textbf{AI-LIT-MH} test was still not predictive of performance or reflective of improvement due to training.

    \begin{mdframed}[backgroundcolor=gray!10, linewidth=0.5pt]
    \textbf{Example question:}\\
    Which of these statements about generative AI is FALSE?\\
    A. It learns from the data provided by humans\\
    B. It has a deep understanding of the content it generates \textbf{(correct answer)}\\
    C. Its outputs can have the same kinds of biases that humans have\\
    D. Its outputs depend heavily on the quality and scope of its training data
    \end{mdframed}

    \item \textbf{Scenario-based instrument (COMP-MCQ):} Finally, we designed a test that we originally intended, and administered, as a competition task, but which we now propose as a new AI literacy test. This test emphasized measurement of practical skills and knowledge that will be required for the role the trainees were preparing for. The results from this test correlated well with competition performance.

    These competition questions differed from the \textbf{AI-LIT-H} and \textbf{AI-LIT-MH} questions in several ways:
\begin{itemize}
    \item \textbf{AI literacy topic alignment} The competition questions in \textbf{COMP-MCQ} were designed to align with the same themes we used when designing the AI User training. This was an analysis of topics we developed through a series of interviews with researchers in AI-Human collaboration, about skills that will be relevant to the future AI-adjacent workforce.  These cross-cut Long and Magurko's categories.
    \item \textbf{Domain Context} The questions were all framed in a naval context, describing scenarios in which navigation, aircraft maintenance, base security, or detection of adversaries might hinge on decisions about appropriate use of AI.  These were designed and reviewed by AI subject matter experts, but not naval experts, and were likely understandable to Navy and Marine participants but not grounded in their actual practices.
    \item \textbf{Bloom's taxonomy level} Bloom observed that learning objectives used in education can vary in depth, ranging from simple memorization up to integrative, creative use of knowledge and skills.  Kraftwohl's revision~\cite{krathwohl_revision_2002} of Bloom's taxonomy classifies cognitive learning objectives into six levels: Remember, Understand, Apply, Analyze, Evaluate, Create.  While the \textbf{AI-LIT-H} and \textbf{AI-LIT-MH} questions were generally at the Remember and Understand level, the \textbf{COMP-MCQ} questions tended to involve application, analysis, and evaluation, by setting up scenarios in which competitors needed to use their understanding of AI systems to analyze a situation and decide what the cause of a problem might be, or what should be done to fix it.
\end{itemize}

    \begin{mdframed}[backgroundcolor=gray!10, linewidth=0.5pt]
    \textbf{Example scenario-based question:}\\
    \textit{Scenario:} You are validating an AI model that predicts landing gear failures. The model’s predictions are being compared to maintenance reports written by engineers. However, after cross-checking the data, you find that many of the engineers' reports contain vague descriptions, such as “mechanical issue” or “general maintenance,” rather than specific failures.\\[0.5em]
    \textbf{Question:} How does this affect the AI model?\\
    A. The AI model will naturally become more accurate over time as it processes more data.\\
    B. The AI model will still function correctly because it learns directly from sensor data, not from maintenance reports .\\
    C. The AI model can compensate for vague data by increasing the number of training examples.\\
    D. The AI model may struggle to learn useful patterns because the labels in the training data are inconsistent. \textbf{(correct answer)}
    \end{mdframed}

\end{itemize}

\section{Context}

\subsection{Carnegie Mellon Robotics Academy Training}
The Carnegie Mellon Robotics Academy (CMRA) 
is working with the US Navy to explore novel training methods for enlisted personnel in robot operation and maintenance roles. The core training consisted of hands-on work in mechatronics, systems integration, and autonomy. By the time participants began the AI-focused module, they had spent at least two weeks working with robots in context.  The Autonomy Foundations module, which overlapped conceptually with AI content, was positioned differently in the two cohorts to manage content sequencing and reduce confounds.

At the conclusion of the six-week training, all participants competed in a series of three full-day events. Each event was designed to test a particular subset of robotics skills through practical application on systems the competition participants had not seen before, in order to test a broader project hypothesis about rapid skill transfer. The competition events were revised between the first and second cohorts; only the second version contained AI-focused competition tasks.

\subsection{AI User modules in context of Robotics training}
\begin{wraptable}{r}{0.55\linewidth}
\centering
\begin{tabular}{l}
\toprule
AI Capabilities and Non-Capabilities\\
The Role and Quality of Data \\
Limitations, Bias, Evaluation and Troubleshooting\\
AI Hardware requirements and workloads \\
* Language Technologies/LLMs\\
Computer Vision \\
* Decision Support and Uncertainty\\
Ethical AI, Data Limitations, Model Fairness \\
\bottomrule
\end{tabular}
\caption{AI User course components.\\ * = omitted for this training}
\label{tab:aiuser-modules}
\end{wraptable}
Robotics use in the Navy, as elsewhere, increasingly involves the use of AI systems. Accordingly, the US Department of Defense has been investigating potential roles and expectations for enlisted personnel who interact with systems that incorporate AI. The Chief Digital and Artificial Intelligence Office (CDAO) has identified relevant AI-related tasks as potentially including data collection and storage, as well as model selection, curation, and evaluation \cite{dod_autonomy_2012}. This does not imply that a robot operator must be capable of, or responsible for, AI model design and validation. On the contrary, one of the objectives of our research is to contribute to the responsible definition and bounding of such activities through empirical determination of which AI-related tasks can demonstrably be performed by personnel without advanced engineering training, and what kinds of training activities or measurable prior knowledge accurately predict success at such tasks. For example, AI literacy could help trainees identify failures in a visual recognition system caused by environmental mismatches in training data, enabling them to collect and submit improved data for retraining.

To address this need, the experimental curriculum included training, assessment, and competition components relating to AI. Six AI User modules were delivered through lecture/discussions and online activities. The modules train students to be informed users of AI and machine learning (ML) based systems.  The modules provide a simulated environment where students practice operating various AI/ML-powered systems in relevant application areas. The modules also discuss the role data plays in AI/ML-powered applications, how to validate and troubleshoot such applications, how to identify when an AI/ML system fails, how to identify inherent biases and issues with fairness within available data and mitigate their effects during decision-making, as well as the impact of computing devices and environments AI/ML systems run on.  Finally, they discuss long-term growing trends in the deployment of AI/ML-enabled automation, AI's current capabilities/limitations, and typical sources of error. 

The AI User units were delivered by beginning with a brief knowledge pre-quiz, then a project consisting of several activities in a simulated environment, followed by a post-quiz (identical to the pre-quiz).  Within each project, the unit presents a ``motivation task'' describing how the learning objectives relate to real-world situations; then several tasks that engage learners in making decisions with and about AI systems. Finally, a transfer task asks students to apply their knowledge to a Navy-specific context.

Table~\ref{tab:aiuser-modules} lists the AI User training's has eight components, derived from a set of interviews we held in 2023 with four experts from one of the NSF's AI Institutes (AI-SDM)~\footnote{\url{https://www.cmu.edu/ai-sdm/}}. Six of the components were used in the training described in this paper; we excluded the language technologies and decision support trainings because they were less relevant to the robotics training.

\section{Methods}
We ran two iterations of the training (Table~\ref{tab:schedule}); the first in June, 2024, and the second in March 2025.

\subsection{Training 1}
The first training involved 13 trainees. Trainees took an AI Literacy pre-test, \textbf{AI-LIT-H}, before the whole training, took self-efficacy instruments for robotics and ML before and after Week 6 (the week of the AI User training), and took single-question pre/post knowledge quiz before and after each of the 6 AI User modules.  Due to a scheduling mishap, only 3 of the 13 took the final AI Literacy post-test, \textbf{AI-LIT-H}.

Additionally, five Marines competed with the graduates of the first cohort as a comparison group, and took the AI Literacy test once.  We analyze the Marines' performance on the \textbf{AI-LIT-H}, but not the rest of their competition results in this paper, since the first competition did not touch on AI topics.

\subsection{Training 2}
For the second training we revised both the robotics and AI User training, as well as made improvements in the assessment instruments. AI User training navigation and logging was improved, and the training was repositioned to align better with the robotics training.  We also increased (from 1 to 5) the number of questions testing knowledge before and after each AI User module, allowing more robust analysis of performance. Figure~\ref{tab:schedule} shows the sequence of activities in both trainings.

Additionally, the second cohort's training was followed up with a competition against a control group of ISR (Intelligence, Surveillance, \& Reconnaissance) specialists near the beginning of their career (0-4 years of service, roughly).  A subset of the competition tasks were designed to mimic practical use of AI in a robotics warfare context.

\begin{table} 
\label{fig:sequence2}
\begin{tabular}{llll}
\textbf{Phase} & \textbf{Date} & \textbf{Trainees} & \textbf{Control} \\ \hline
Training 1 & July 2024 & \begin{tabular}[c]{@{}l@{}}Pre \textbf{AI-LIT-H} (13)\\ Robotics training\\ 
AI user training\\ 
Post \textbf{AI-LIT-H} (12)\end{tabular} &  \\ \hline
Competition 1 & July 2024 & Competition tasks (5) & \begin{tabular}[c]{@{}l@{}}
\textbf{AI-LIT-H} (5) \\
Competition tasks (5)
\end{tabular} \\ \hline
Training 2 & Jan/Feb 2025 & \begin{tabular}[c]{@{}l@{}}Pre-\textbf{AI-LIT-MH} (13)\\ 
Robotics training\\ AI User training\\ 
Post-\textbf{AI-LIT-MH} (13)\end{tabular} &  \\ \hline
Competition 2 & March 2025 & Competition tasks (5) & \begin{tabular}[c]{@{}l@{}}
\textbf{AI-LIT-MH} (5)\\ Competition tasks (5)\end{tabular}
\end{tabular}
\caption{Schedule of tests, trainings, and competitions. Number of completed data items in parentheses. Note that one of the competition tasks is  \textbf{COMP-MCQ}.}
\label{tab:schedule}

\end{table}  

\subsection{Measurements}

\subsubsection{AI Literacy test}
Our goal was to develop an AI Literacy test intended to assess the competencies required by non-IT professionals for effective interaction with AI tools in real-world, task-oriented environments. It is meant to gauge how well individuals can use AI systems to enhance their job performance, make informed decisions, and troubleshoot when issues arise.

\subsubsection{Demographics}
We asked participants to report their age, gender, and highest level of education completed. For the second cohort, we added open-ended questions to better understand their prior experience, since their preparation for the course varied widely. We asked whether participants had used AI-powered tools (and if so, which ones and how), whether they had ever disagreed with an AI system (with examples), what their programming background was, and whether they had participated in any AI-related competitions or "AI hacking" activities.

We manually converted the education level scores to a ranking from 1-4 (high school, some college, undergrad degree, or advanced degree), and similarly created an experience ranking by counting how many answers indicated some experience.

\subsubsection{Cohort 2 Competition Tasks}
We hypothesized that the AI User training would improve students' scores in tasks that depended on knowledge of how to interact with AI systems (e.g., capturing and labeling an appropriate set of images to train a navigation model).

The competition had four tasks that were relevant to AI literacy:
\begin{itemize}
    \item \textbf{Dataset (\textbf{COMP-DATASET})} (Capture and Organize): Competitors were given a detailed scenario involving the deployment of autonomous ground robots to detect intruders. They were tasked with collecting a dataset of images using a smartphone camera, simulating the perspective and constraints of the robots. Participants then organized the images on a provided laptop and created a spreadsheet with descriptive metadata for each image, simulating the process of preparing training data for an object detection algorithm.
    \item \textbf{Reflection (\textbf{COMP-OPEN})}: Participants answered a series of open-ended questions designed to elicit their reasoning process, the challenges they encountered while collecting and annotating the dataset, and their understanding of the practical use of AI in the given scenario.
    \item \textbf{Multiple Choice Questions (\textbf{COMP-MCQ})}: Participants completed a set of scenario-based MCQs testing their ability to apply AI concepts to realistic dilemmas, particularly in operational and military contexts. These questions assessed judgment, trade-off reasoning, and practical AI knowledge. This is the competition task that we propose to repurpose as an AI literacy instrument.
    
    \item \textbf{Hammerhead Model Peak F1 (\textbf{COMP-HH})}: A quantitative measure of participants' performance in selecting and labeling data to improve AI model performance. Participants made 3-minute robot reconnaissance ``runs'' in an enclosed space containing various objects of interest (e.g., footballs). Each participant used an onboard camera to capture images during these runs, then curated and labeled their images, and used them to train ``custom'' classes in a YOLO model to automatically identify the objects (a contingency dataset of usable images was also made available in case participants failed to capture good images themselves in the first portion).

    We report ``Peak F1'' performance on the classifier-training portion of this task---the most-favorable F1 score a participant's model obtained under a range of thresholds, i.e., the best balance of true vs. false object reports, under the most generous interpretation of each model.
\end{itemize}

Note that there were other, robotics-focused competition tasks, not described here.

\section{Results}

\begin{wrapfigure}{r}{0.5\linewidth}
    \centering
    \includegraphics[width=0.45\linewidth]{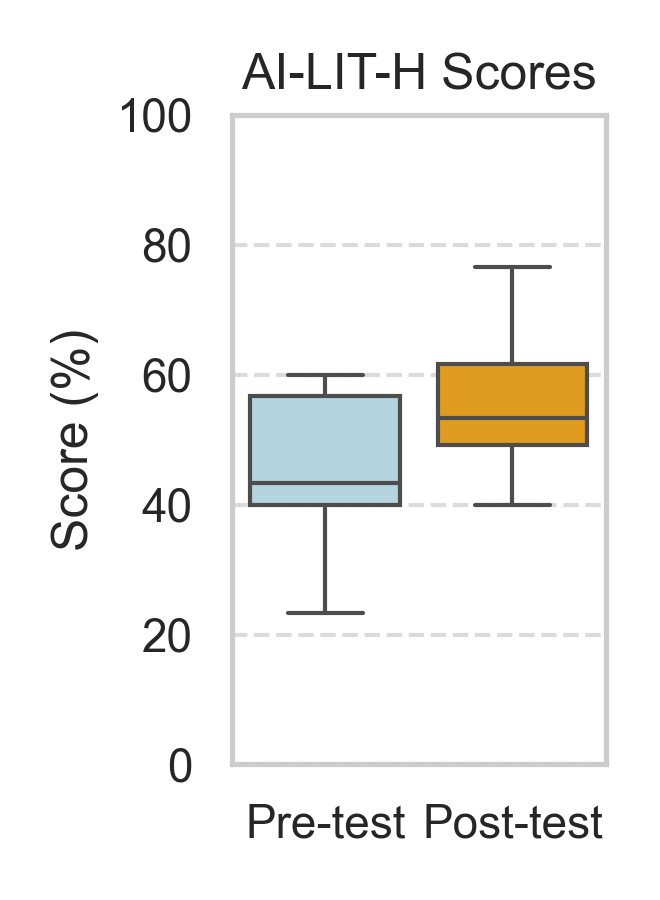}
    \includegraphics[width=0.45\linewidth]{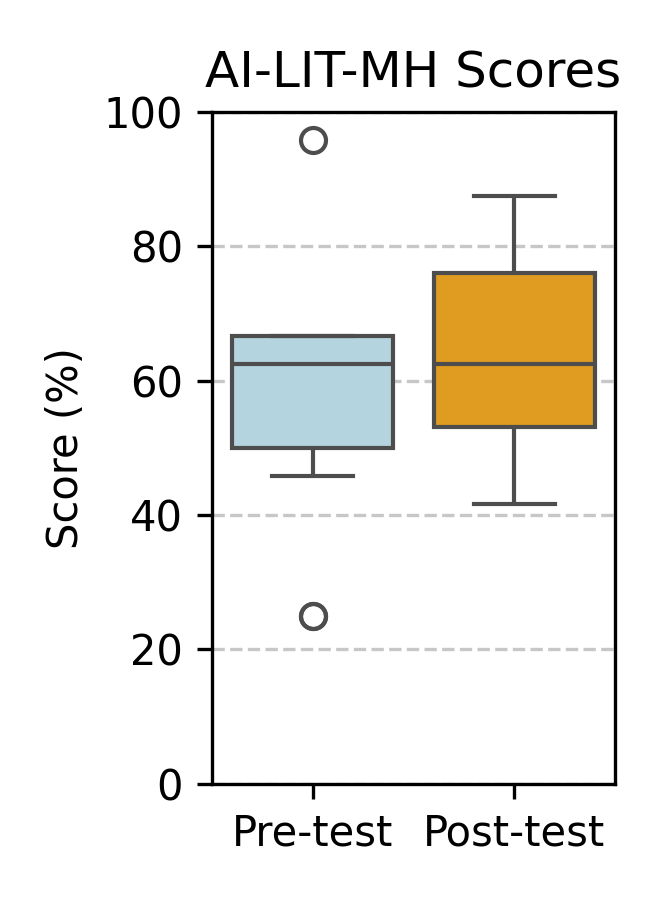}
    \caption{Neither \textbf{AI-LIT-H} nor \textbf{AI-LIT-MH} increased after AI User training.}
    \label{fig:no-ai-lit-gains}
\end{wrapfigure}
Although the context of this work is a robotics training and competition, we focus our analysis on AI Literacy: the measurement instruments, the training, and AI Literacy's interaction with performance on other tasks.  The structure of the competition means different data is available for different participants: for example we do not have pre-post measurements for the competition control group, since they only took \textbf{AI-LIT-MH} 
once, before the competition.

Across both cohorts, participants showed significant learning gains  on post-unit quizzes (e.g., p=0.0002 in Cohort 2), but no significant improvement in AI Literacy test scores (\textbf{AI-LIT-H}: p=0.165; \textbf{AI-LIT-MH}: p=0.28). (Figure~\ref{fig:no-ai-lit-gains})

\subsection{Results: First CMRA cohort}

The first cohort's trainees were all male, five from the Navy and eight recruited from local community colleges. The median age was about 24 years old. Eight had a high school education, three had associates degrees, and one had a bachelor's degree.

The competition in the first cohort did not have an explicit AI component, so we only use the competition control group as additional data points for \textbf{AI-LIT-H}.

\subsubsection{AI Literacy instrument}
\label{sec:ailit_hornberger}
Our item analysis of the \textbf{AI-LIT-H} \cite{hornberger_what_2023} revealed several ways in which it did not fit the work-task focused notion of AI Literacy.

\emph{Outdated information.} Several questions reflect assumptions that no longer hold. For instance, one item suggested that humans are superior to AI in proving mathematical theorems, but AI systems have since demonstrated success in this area. Another claimed that irony is better understood by humans than AI—a distinction that no longer reliably applies with modern language models.

\emph{Limited workplace relevance.} Some items test esoteric knowledge with little bearing on practical use of AI. For example, distinguishing whether a field like NLP is a subfield of AI may be challenging or irrelevant for non-experts, especially when contrasted with distractors like bioinformatics or psychology, that may be unfamiliar to the test-taker.

\emph{Debatable or opinion-based claims.} A few questions rely on philosophical or loosely defined assertions, such as equating human learning with machine learning. These are open to interpretation and not reliably answerable in a workplace context.

\emph{Missing topics.} The test also omits key areas of modern AI practice, such as generative AI, model evaluation, and responsible usage—topics that are essential for informed AI use in real-world settings.

\subsection{Results: Second CMRA cohort}

The second cohort had thirteen trainees, and five competitors brought in after the training, for a total of 18, all male.  Sixteen had a high school education, one had a bachelor's degree, and one had not completed high school.  The trainees' average age was 27, ranging from 22 to 40, and the control group ranged from 19-25.

\subsubsection{\textbf{AI-LIT-MH} item analysis}
\label{sec:ailit_item_analysis}
We performed several item analyses on \textbf{AI-LIT-MH}, including mean score for each question, discrimination ability, and correlation with total score. However because our population was  small, we treated the results as qualitative indicators that something might be wrong with either the training or the test, rather than removing items simply because of their scores.

We removed two questions about the limitations of AI.  One was poorly correlated with the test and students got this question wrong more often after the training than before: the question relied on a definition of Weak and Strong AI which we borrowed from \textbf{AI-LIT-H}, but in fact it was not consistent with how the term has been used.  We also removed a second question in the same section that also relied on this poor definition, although participants did get it right most of the time.

We also removed one question from the Outputs of AI section: participants scored worse on it after completing AI User training. Participants' most common ``wrong'' answer seemed plausible to us, opining that the worst problem with black box AI is that people may trust their decisions without question.  We thus discarded it.

Four other items that stood out as having low correlation with the overall test score; two in data literacy, one in Tool2Task, and one in Ethics. These all related to facts about AI training and ethics that we believed were important, but that participants who otherwise scored well on the instrument simply didn't always know.  We opted to keep these in the test on the grounds that the questions seemed to have good face validity.

After removing these items, we analyzed how their subscores had changed. The results show our training's alignment with the test was best for Tool2Task (i.e. appropriate uses of AI tools), and the limitations of AI.  

The least alignment was in the Data Literacy category, where students did significantly worse after the course.  These questions related to the details of manipulating datasets when training AI tools.  The training does incorporate practice with these skills, but does not emphasize terminology.

\subsection{Results: Second cohort competition}

Five ISR specialists were brought in at the end of the training for a competition to demonstrate the efficacy of the training.  The competition results will be reported in full elsewhere, but here we use the additional participants and competition tasks to gather further data about AI Literacy.

\begin{table}

\begin{tabular}{lccc}
\toprule
 & \textbf{AI-LIT-MH} & \textbf{AI User quiz scores} & \textbf{COMP-MCQ} \\
\midrule
\textbf{COMP-MCQ} & \cellcolor{green!20} 0.74 & \cellcolor{green!20} 0.80 & \cellcolor{green!20} --- \\
\textbf{COMP-OPEN} & 0.24 & 0.29 & \cellcolor{green!20} 0.70 \\
\textbf{COMP-DATASET} & 0.51 & 0.42 & 0.39 \\
\textbf{COMP-HH} & 0.43 & 0.44 & 0.61 \\
\bottomrule
\end{tabular}

\caption{Competition subscore correlations with quiz results for the \textbf{AI-LIT-MH} score, the AI User portion of the training, and one of the competition subscores again. Using \textbf{COMP-MCQ} as an AI literacy measure predicts two of the competition tasks better than \textbf{AI-LIT-MH} did. n=10 participants with all 6 measures.}
\label{tab:score-corr-with-competition}
\end{table}

As shown in Table~\ref{tab:score-corr-with-competition}, \textbf{COMP-MCQ} was sometimes more predictive of applied task performance than \textbf{AI-LIT-MH}, significantly correlating with two of the competition tasks. \textbf{AI-LIT-MH} was weakly related to the dataset task, but less predictive overall.

\subsection{Results: Scenario-based assessment}
Table~\ref{tab:score-corr-with-competition} shows how \textbf{COMP-MCQ} correlates both with other competition tasks, and with \textbf{AI-LIT-MH}. \textbf{COMP-MCQ} correlates significantly with two of the other competition tasks, and also with performance on the quizzes given after the AI User training.
The correlations between this competition score and the other competition scores is higher for two of the tasks than correlations with \textbf{AI-LIT-MH} or its subscales.  However for the \textbf{COMP-OPEN} task asking for open-ended explanations of the competitor's choices, the more knowledge-based \textbf{AI-LIT-MH} was more predictive.

\begin{figure}[hbt]
    \centering
    \includegraphics[width=0.45\linewidth]{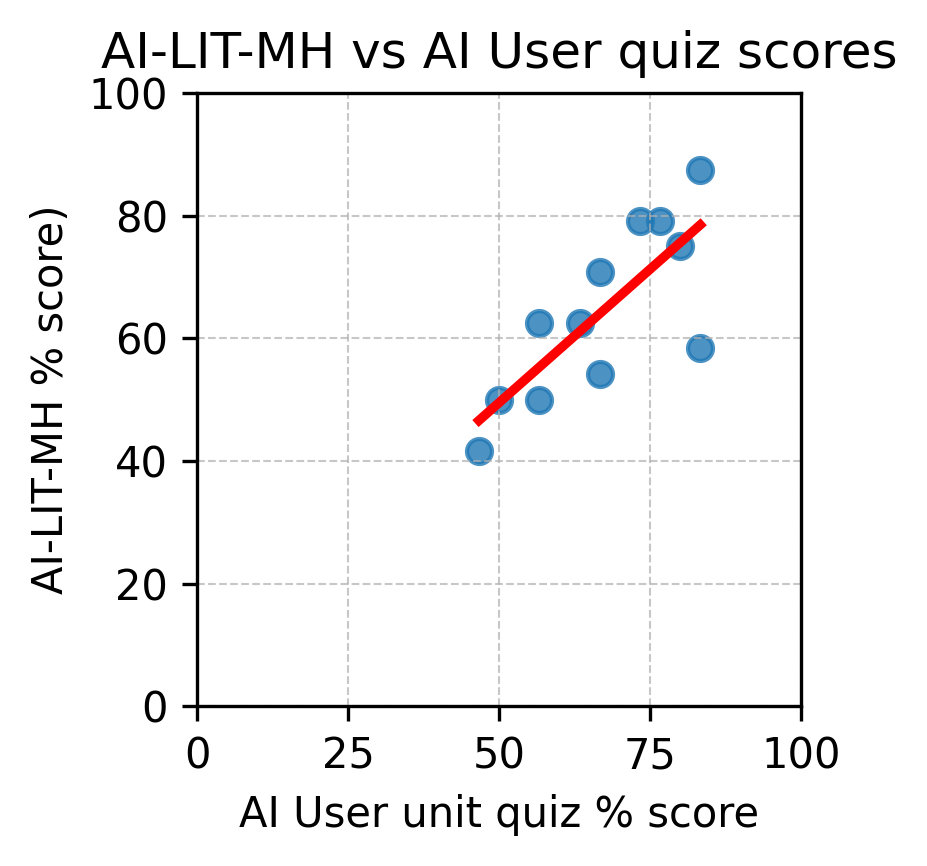}
    \includegraphics[width=0.46\linewidth]{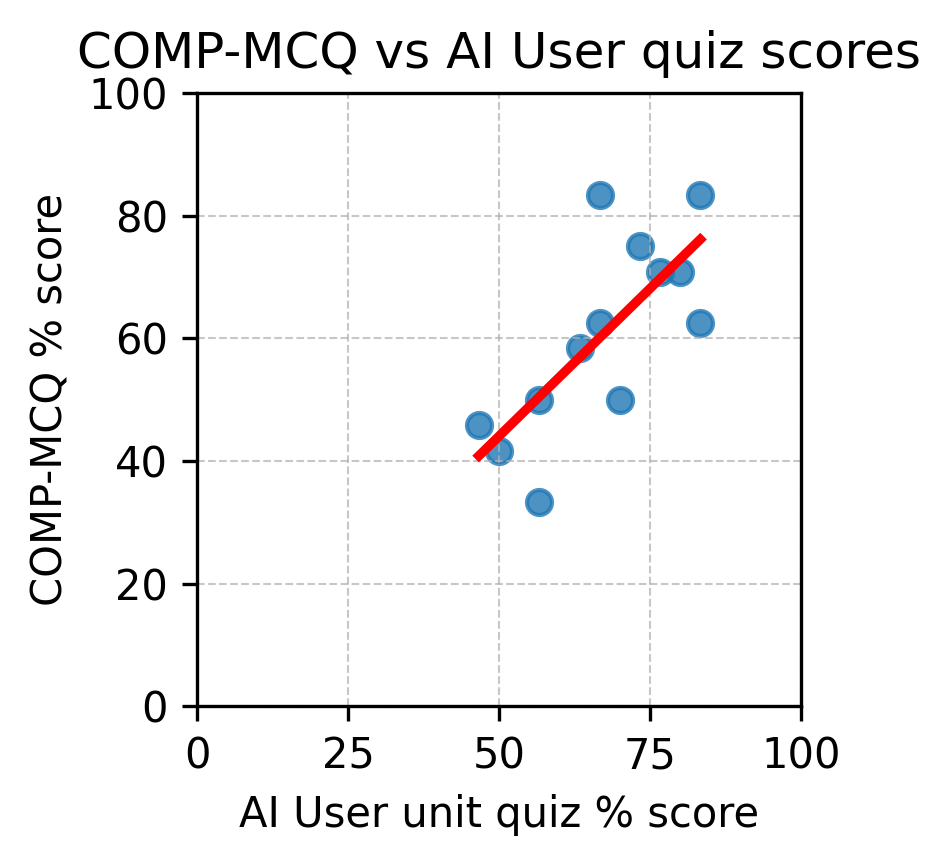}
    \caption{Both \textbf{AI-LIT-MH} and \textbf{COMP-MCQ} instruments had a significant relationship with performance on the AI User training.}
    \label{fig:regressions}
\end{figure}

Both instruments showed significant relationships with AI User quiz scores (Figure~\ref{fig:regressions}; p=0.003 and p=0.004, respectively).  Note that while both measures AI literacy correlated with final AI user quiz scores, \textbf{AI-LIT-MH} did not appear to capture AI User learning (see Figure~\ref{fig:no-ai-lit-gains}). (We could not assess \textbf{COMP-MCQ}'s capture of learning, since it was only administered after the AI User training).

\section{Discussion and Future Work}

We struggled to make off-the-shelf AI Literacy instruments serve the purpose of measuring learning from our training, or to predict success in a practical task.  This suggests that current instruments may be mismatched with the goals of modern AI literacy efforts, particularly those aimed at non-technical professionals. To prepare workers to make informed decisions about AI use, we will need assessments that reflect as closely as possible the judgments that people in the field will actually be expected to make, regardless of the theory or terminology behind such judgments. In our particular case, this  will require incorporating scenario-based items grounded in real work tasks. 

Our own future work will involve building off these scenario questions and exploring whether it was the \emph{domain} or the \emph{Bloom level} that mattered: If we had situated the scenarios in non-naval domains, would the assessment have worked as well?  On the other hand whether more vetting by naval subject matter experts would have provided even better alignment with task performance. For example we could have incorporated realistic situations Navy personnel might encounter.

For our team this will be important to resolve as we offer our training in other domains.  We intend to do some specialization of the training to other work domains, such as public health and disaster response, and the degree to which AI literacy evaluation needs to be specialized will affect the cost of our efforts.

\section{Conclusion}
Our findings suggest that AI literacy, as currently measured by widely used instruments, may not align well with the practical competencies required for real-world decision-making with AI. In contrast, scenario-based assessments, involving analysis, application, and evaluation of knowledge in authentic, domain-relevant tasks may better assess the applied knowledge and judgment that workforce-oriented training aims to cultivate. Rethinking the way we evaluate AI training for the workforce invites an alternate framing of AI literacy itself not as a static body of knowledge but as a situated capacity to act responsibly and effectively in any given domain.

\section{Acknowledgments}
Notice: This work relates to Department of Navy award \#N0001423C2015 issued by the Office of Naval Research. The United States 
Government has a royalty-free license throughout the world in all copyrightable material contained herein. Any opinions, findings, and conclusions or recommendations expressed in this material are those of the author(s) and do not necessarily reflect the views of the Office of Naval Research.

Hosting of the educational platform was generously supported by Microsoft. 

\bibliography{main}

\end{document}